\documentstyle [preprint,float,aps,psfig]{revtex}
\tightenlines

\newcommand{\inqq} {\int \! {\frac {d^3 \vec q} {(2\pi)^3}}
  \frac {d^3 \vec q_1}{(2\pi)^3}\,\psi_d (\vec q)
  \psi_d (\vec q _1 -\frac {\vec \Delta}{2})}

\begin{document}

\title{Charge Symmetry Violation Effects in Pion
Scattering off the Deuteron}

\author{V.V. Baru\thanks{baru@vxitep.itep.ru}, 
        A.E. Kudryavtsev\thanks{kudryavtsev@vitep5.itep.ru},
    and V.E. Tarasov\thanks{tarasov@vitep5.itep.ru}}
\address{Institute Theoretical and Experimental Physics \\
25 B. Cheremushkinskaya Street, Moscow, Russia 117259}

\author{W.J. Briscoe\thanks{briscoe@gwu.edu}, 
        K.S. Dhuga\thanks{dhuga@gwu.edu}, and 
        I.I. Strakovsky\thanks{igor@gwu.edu}}
\address{Center for Nuclear Studies and Department of 
Physics, \\
The George Washington University, Washington, DC 20052}

\draft

\date{\today}
\maketitle

%%%%%%%%%%%%%%%%%%%%%%%%%%%%%%%%%%%%%%%%%%%%%%%%%%%%%%%%%%%%
\begin{abstract}
We discuss the theoretical and experimental situations for
charge symmetry violation (CSV) effects in the elastic
scattering of $\pi ^+$ and $\pi ^-$ on deuterium (D) and 
$^3He/ ^3H$.  Accurate comparison of data for both types 
of targets provides evidence for the presence of CSV 
effects.  While there are indications of a CSV effect in 
deuterium, it is much more pronounced in the case of 
$^3He/ ^3H$.  We provide a description of the CSV effect 
on the deuteron in terms of single- and double-scattering 
amplitudes.  The $\Delta$-mass splitting is taken into 
account.  Theoretical predictions are compared with 
existing experimental data for $\pi-d$ scattering; a 
future article will speak to the $\pi$-three nucleon 
case. \\ 
\end{abstract}

%%%%%%%%%%%%%%%%%%%%%%%%%%%%%%%%%%%%%%%%%%%%%%%%%%%%%%%%%%%%

\narrowtext
\section{Introduction}
\label{sec:intro}

The study of CSV in the interaction of pions with nuclei 
in the Delta resonance region has been of considerable 
interest for the last two decades.  The interaction of 
pions with light nuclei such as $^2H$ \cite{pe78} $-$ 
\cite{ne90}, $^3He/ ^3H$ \cite{ne90} $-$ \cite{br99}, and 
$^4He$ \cite{br91} has attracted particular attention.  
However, we note that quite a large data set also exists 
for scattering of $\pi ^+$ and $\pi ^-$ on $^{12}C$, 
$^{16}O$, and $^{40}Ca$ as well \cite{dh99}.

 From the point of view of theory, the advantage of 
searching for CSV in the scattering of pions from light 
nuclei is that one can describe pion scattering in these 
systems in a relatively straight-forward manner.  With 
this in mind, we limit ourselves to the consideration of 
the scattering of pions from deuterium, $^3He$, and $^3H$.  
Moreover, we anticipate that CSV effects are considerably 
diminished in the case of pion scattering from heavier 
nuclei because of the importance of processes such as 
absorption.  

First, in order to evaluate the scale of CSV effect, 
we focus our theoretical efforts primarily on $\pi d$ 
elastic scattering.  In a following article, we will 
develop the formalism further to investigate CSV in 
the three-nucleon system.

A detailed analysis of the experimental situation will 
be given in the next section.  Here, we want only to 
point out that in order to make a comparison between 
experimental data related to different projectile or 
target, we must deal with the same experimental 
measurables.  Historically, the CSV experimental data 
were given in terms of asymmetry, $A_{\pi}$ for the 
deuteron:
\begin{eqnarray}
A_{\pi} = \frac
{d \sigma /d \Omega ( \pi ^- d) - d \sigma /d 
\Omega ( \pi ^+ d)}{d \sigma /d \Omega ( \pi ^- 
d) + d \sigma /d \Omega ( \pi ^+ d)}, 
\label{1} \end{eqnarray}
and in terms of ratios $r_1$ and $r_2$, and superratio 
$R$ for the $^3He/ ^3H$ case:
\begin{eqnarray}
r_1 = \frac
{d \sigma /d \Omega ({\pi ^+} {^3 H})}
{d \sigma /d \Omega ({\pi ^-} {^3 He})}, 
\phantom{xxxxxxxx} \nonumber \\
r_2 = \frac
{d \sigma /d \Omega ({\pi ^-} {^3 H})}
{d \sigma /d \Omega ({\pi ^+} {^3 He})}, 
\phantom{xxxxxxxx} \nonumber \\
R = r_1~r_2. \phantom{xxxxxxxxxxxxxxxxx}
\label{2} \end{eqnarray}
Both interactions ${\pi ^+}{^3H}$ and ${\pi ^-}{^3He}$ 
for the ratio $r_1$, and ${\pi ^-}{^3H}$ and ${\pi ^+}
{^3He}$ for the ratio $r_2$ are isomirror interactions.  
Therefore, if charge symmetry is strictly observed, 
both $r_1$ and $r_2$ would be equal to $1.0$.  Of 
course, the Coulomb interaction is not charge symmetric 
and would have to be taken into account.  The 
superratio $R$ is the product $r_1$ and $r_2$.  So, if 
charge symmetry is universally true, $R$ is also equal 
to $1.0$.  

The experimental data suggests evidence for a small 
effect in $A_{\pi}$ for the deuteron (e.~g. $A_{\pi} 
\simeq 2\%$ at 143~MeV \cite{ma82}) with some 
indication of structure at scattering angles around 
$90^{\circ}$ in cm frame.  At the same time, a sizable 
effect is clearly seen in the $^3He/ ^3H$ case.  For 
example, $r_2 = 0.7 \pm 0.1$ for $T_{\pi}$ = 256~MeV 
and $\theta = 82^{\circ}$ \cite{dh96}.  Theoretical 
predictions for the asymmetry $A_{\pi}$ in the 
deuterium case were given in Ref. \cite{ma82}.  To 
describe the asymmetry, authors of Ref. \cite{ma82} 
used a single-scattering approximation with allowance 
for differently charged $\Delta$'s(1232).  In this 
approximation, the CSV effect proved to be 
independent of the scattering angle with typical 
value proportional to $\delta m_{\Delta} / \Gamma 
_{\Delta}$.  Approximately the same approach was 
used in the $^3He/ ^3H$ case in Ref.\cite{ne90}.

A different approach for the $^3He/ ^3H$ case was 
suggested in the paper \cite{ki87}.  Authors of this 
paper used an optical potential to describe the 
pionic $^3He/ ^3H$-amplitudes.  The radial 
dependence of $\pi A$ potentials was determined in 
terms of matter and spin densities for $^3He$ and 
$^3H$.  The Coulomb-nuclei interference effect in 
the vicinity of minima in differential cross 
sections was reported as the main reason for the 
CSV effect in \cite{ki87} approach.  However, this 
interpretation was disputed by Briscoe and 
Silverman \cite{br89} because the authors of 
\cite{ki87} obtained structure only near the 
$90^{\circ}$ in $r_2$ but could not at all explain 
the overall behavior of the experimental data.

In our investigation, we study the role of 
double-scattering on CSV because of mass splitting 
of $\Delta$-isobars.  It is widely known that the 
single-scattering approximation reproduces a 
differential cross section fairly well in the 
forward hemispere.  But for scattering angles 
beyond $90^{\circ}$, the double-scattering term is 
important and should be included.  The influence of 
multiple scattering terms on differential cross 
section for deutron case was studied long ago in 
the papers \cite{Gibbs} $-$ \cite{Kudr2}.  But the 
influence of double and multiple scattering on CSV 
effects was never studied in detail.

In Section~III, we explain how the basic ingredients 
of the scattering amplitude and constraints such as 
single- and double-scattering, and the Coulomb 
interaction are combined for $\pi d$ elastic 
scattering.  These results and the prospect for 
improvement are summarized in Section~IV.  The 
$^3He/ ^3H$ case is considered in forcoming paper.

\section{Analysis of Experimental Situation}
\label{sec:exp}

The CSV effect was first observed in the difference 
of total $\pi ^{\pm} d$ cross sections in PSI and
reported in \cite{pe78}.  This has been widely 
discussed, see, e.~g. the book by Ericson and Weise 
\cite{EW}.  There have been several measurements for 
both $\pi ^+ d$ and $\pi ^- d$.  The first systematic 
study of the CSV effect in the differential $\pi ^{\pm} 
d$ cross sections was done at LAMPF and presented in 
the paper \cite{ma81}.  Soon after, the asymmetry 
$A_{\pi}$ for $T_{\pi}$ = 143~MeV was presented for 
the range of laboratory scattering angles between 
$20^{\circ}$ and $115^{\circ}$ \cite{ma82}.  The 
experiment was repeated for approximately the same 
range of scattering angles at $T_{\pi}$ = 256~MeV 
\cite{ma84}.  We note that the structure in the 
asymmetry seen in \cite{ma82} was not seen in the 
TRIUMF measurements of \cite{sm88}.  Meantime, some 
indications for CSV effects were also obtained at 
low energies 30, 50, and 65~MeV at TRIUMF 
\cite{ko91,ko93}.  We also mention the high-energy 
Gatchina data at $T_{\pi}$ = 417~MeV \cite{gr95}
which also shows some indications on CSV.

We recall that the asymmetry (\ref{1}), and ratios 
(\ref{2}), are two different measures of CSV-effects.  
As in the $^3He/ ^3H$-case, we denote the ratio $r = 
r_1 = r_2$ $$r = \frac{d \sigma /d \Omega ( \pi ^- 
d)}{d \sigma /d \Omega (\pi ^+ d)} = 1 + \epsilon .$$  
Then, in the case of small magnitudes of CSV, we get 
$$A_{\pi} \approx \epsilon /2.$$  Clearly, this tiny 
effect would require high-quality data.  

Smith {\em {et al.}} \cite{sm88} reported a $-$1.5\% 
asymmetry in the $\pi d$ cross sections at back angles, 
with uncertainties of 0.6\% at the different angles.
The energy dependence of the asymmetry between 30 and
417~MeV is shown in Fig.~1.

\section{Theoretical Consideration of CSV-effect 
in Deuteron}
\label{sec:theor}

We see two possible ways to interpret the 
experimental situation:

\begin{itemize}
\item
The {\it first} way is that one may conclude that 
there is really no effect in deuterium in accordance 
with statement \cite{sm88} and that the effect in 
the $^3He/ ^3H$ case is influenced correspondingly 
by specific three-body configurations of $^3He$ and 
$^3H$.  By this, we mean the possible influence of  
three-body, CSV forces which are absent in the $^2H$ 
case and/or differences in the description of $^3He$ 
and $^3H$ wave functions (WF) as a consequence of 
an additional Coulomb repulsion between two protons 
in the $^3He$ case (see in this connection Ref. 
\cite{GG}). 

\item
The {\it second} scenario is to suggest that the 
effect may be seen in both cases $^2H$ and $^3He/ 
^3H$, but in deuterium, the effect is small in 
comparison with  $^3He/ ^3H$.  There should 
still be some angular dependence for the CSV effect 
in deuterium.  However, Masterson {\em {et al.}} 
\cite{ma82} have shown that within the impulse 
single-scattering approximation the angular 
dependence for CSV is absent when only scattering 
via the $P_{33}$ is considered.  The inclusion of 
others S- and P-waves does not change the situation 
dramatically as all the phases except $P_{33}$ are 
small in the region of interest.  So, we need to 
look beyond the single-scattering approximation and 
to consider multiple scattering of pions.
\end{itemize}

\begin{enumerate}
\item
{\underline {Single-Scattering Approximation}}

Everywhere below, we shall use the following 
notations: $k_{cm} = \frac{m}{m + \omega} k$, $w
= m + \omega - \frac{k^2}{2 (m + \omega)}$, where 
$\omega$ is the pion energy, $w_i$ are the masses 
of isobars, and here and below indices 
$1-4$ in the notations of amplitudes, masses and 
widths mean the corresponding isobar isospin 
state: $$i = 1,~2,~3,~4$$ for $$\Delta ^{++},~ 
\Delta ^{+},~ \Delta ^{0},~ \Delta ^{-}.$$
We suppose $\Gamma _{el} = \Gamma _{tot} = \Gamma 
_0 = 120~MeV$.  The values $w_i~(i = 1,~2,~3,~4)$, 
we calculate according to the formula from Ref. 
\cite{EW} (page~124, Eq.~(4.16)): $$w_i = a - b~I_i 
+ c~I^2_i,$$ where $I_i$ is the 3$^{rd}$ component 
of isospin for the $i^{th}$-term from the 
$\Delta$-multiplet.  Using the average resonance 
value from the PDG \cite{PDG} $w_0 = 1232~MeV$, we 
get $a = 1231.8~MeV$, $b = 1.38~MeV$, and $c = 
0.13~MeV$.  In this approximation, the $\pi d$ 
amplitude is the sum of the two Feynman diagrams 
shown in Fig.~2.  

The elementary $\pi N$ amplitude in terms of 
$\delta _{33} (k)$ phase looks like the following:
\begin{eqnarray}
\hat f_{\pi N} = \frac{1}{2 i k}~(e^ {2i \delta 
_{33} (k)} - 1)~\frac{2 + \vec{t}\cdot\vec{\tau}}
{3}~(2~\hat{\vec{k}}\cdot\hat{\vec{k^{\prime}}} + 
i~\vec{\sigma}\cdot [ \hat{\vec{k}}\times 
{\vec{k^{\prime}}}]), 
\label{3} \end{eqnarray}
where $\vec{\sigma}$ and $\vec{\tau}$ are Pauli 
matrices and $\hat f_{\pi N}$ is the operator in 
spin and isospin space of the $\pi N$ system.  The 
deuteron wave function in $S$-wave approximation is 
$$\frac{1}{\sqrt 2}~\psi _d(p)~w^{+}_2~( 
\vec{\epsilon}\cdot\vec{\sigma})~\sigma_2~w^{*}_1$$ 
(here $w_1$ and $w_2$ are the nucleon spinors and 
$\vec{\epsilon}$ is the polarization vector of 
deuteron), and the expression for amplitude $f_1$, 
which correspond to the diagram Fig.~2a, has the 
form:
\begin{eqnarray}
f_{\pi d}^{(1)} = \frac {2}{E^{\pi d}_{cm}}\int 
\frac{d\vec{p}}{(2\pi)^3} E^{\pi N}_{cm} f_{33}
(k_{cm}) \psi_d (\vec {p}) \psi_d (\vec{p} - \frac 
{\vec{\Delta}}{2}) \left (2(\vec{\epsilon}\cdot
\vec{\epsilon^{\prime}}) (\hat{\vec{k}}\cdot\hat
{\vec{k^{\prime}}}) - [\vec{\epsilon}\times\vec
{\epsilon^{\prime}}] \cdot [\hat{\vec{k}}\times
\hat{\vec{k^{\prime}}}] \right) .
\label{4} \end{eqnarray}
Here $\vec{\Delta} = \vec{k} - \vec{k^{\prime}}$ 
is the 3-dimension momentum transfer; $f_{33}(k) 
= \frac{1}{2ik}~(e^{2 i \delta_{33} (k)} - 1)$; 
$\vec{\epsilon}(\vec{\epsilon^{\prime}})$ is the 
polarization vector of initial (final) deuteron; 
$\hat{\vec{k}} = \vec{k}_{cm} / k_{cm}$ and
$\hat{\vec{k^{\prime}}} = 
\vec{k^{\prime}}_{cm}/k_{cm}$ are the units 
vectors, where $\vec{k}_{cm} 
(\vec{k^{\prime}}_{cm})$ is the momentum of 
initial (final) pion in the rest frame of 
subprocess $\pi N\to \pi N$.

At this stage, we make some simplifications.  We 
shall neglect Fermi motion of the nucleon and 
consider (for a while) the expression (\ref{4}) 
in the static limit, i.~e. $\omega/m \to 0$.  
Then, $2 E^{\pi N}_{cm} /E^{\pi d}_{cm} \to 1$, 
$k_{cm} \to k$.  So, we get 
\begin{eqnarray}
\hat f_{\pi d}^{(1)} = \frac{4}{3} f_{33} (k)
\left(2(\vec{\epsilon}\cdot\vec{\epsilon 
^{\prime}}) (\hat{\vec{k}}\cdot 
\hat{\vec{k^{\prime}}}) - 
[\vec{\epsilon}\times\vec{\epsilon^{\prime}}] 
\cdot [\hat{\vec{k}}\times\hat{\vec{k^{\prime}}}] 
\right) \int\Psi _D^2 (r) e^{i\frac{\vec{\Delta}}
{2} \cdot\vec{r}} d \vec r .
\label{5} \end{eqnarray}
For this amplitude, the differential cross section 
with the unpolarized initial deuteron has the 
following form:
\begin{eqnarray}
\frac {d \sigma _{\pi d} ^{(1)}}{d \Omega} = 
\frac{32}{27}~(6 \cos ^2 \theta + \sin ^2 
\theta ) ~| f_{33} (k) | ^2~ F^2_D ( \Delta ),
\label{6} \end{eqnarray}
where $F_D ( \Delta ) = \int \Psi_D^2 (r) 
~e^{\frac{i \vec{\Delta}\cdot\vec {r}}{2}} d 
\vec r$.  This expression agrees with that 
given in Ref. \cite{ma82}.  The ratio 6:1 
between the terms proportional to $\cos ^2 
\theta$ and $\sin ^2 \theta$ reflects the ratio 
of non-spin-flip to spin-flip amplitudes in this 
approximation.  

\item
{\underline {Charge Symmetry Breaking Effect}}

First consider the elementary $\pi ^+ p$ amplitude 
in terms of a $\Delta$(1232) pole.  The amplitude 
looks like a standard Breit-Wigner amplitude
\begin{eqnarray}
f _{\pi ^+ p} = - \frac{1}{2 k}~\frac{\Gamma _{1}}
{w - w _{1} + i~ \Gamma _{1}/2},
\label{7} \end{eqnarray}
where $w_{1}$ and $\Gamma _{1}$ are the mass and  
the full width, respectively, of the $\Delta 
^{++}$ resonance.  Making a linear expansion of 
this amplitude around the mean value of the mass 
$w_0$ and the width $\Gamma_0$ for the $\Delta$ 
resonance, we get
\begin{eqnarray}
f _{\pi ^+ p} = - \frac{1}{2 k}~\frac{\Gamma_0}{w - 
w_0 + i~ \Gamma_0/2}~(1 + \frac{\delta \Gamma _{1}}
{\Gamma_0} + \frac {\delta w_{1} - i~ \delta\Gamma 
_{1}/2}{w - w_0 + i~ \Gamma_0 /2}), 
\label{8} \end{eqnarray}
where $\delta \Gamma _{1} = \Gamma _{1} - \Gamma_0$ 
and $\delta w_{1} = w_{1} - w_0$.  So, using 
Eq.~(\ref{8}), we get that the charge asymmetry in 
$\pi ^{\pm} d$ scattering in this approximation is
\begin{eqnarray}
A_{\pi} = \frac{3}{4}~\frac{C_{\Gamma} (w - w_0) ^2 
+ (w - w_0)~C_M \Gamma _0}{\Gamma_0 ~[(w - w_0) ^2 
+ \Gamma _0 ^2 /4]},
\label{9} \end{eqnarray}
where the parameters $C_M$ and $C_{\Gamma}$ are 
expressed in terms of $\Delta$ mass and width 
splitting:
\begin{eqnarray}
C_M = \delta w_4 + \frac{1}{3}~\delta w_3 - 
\frac{1}{3}~\delta m_2 - \delta m_1 \simeq 
4.6~MeV, \nonumber\\
C_{\Gamma} = \delta\Gamma _4 +  \frac{1}{3}~
\delta\Gamma _3 - \frac{1}{3}~\delta\Gamma _2 - 
\delta\Gamma _1 \simeq 1.7~MeV. 
\phantom{xx} \nonumber
\end{eqnarray}

These values are taken from the Masterson {\em 
{et al.}} paper \cite{ma82} and are in agreement 
with the most recent data \cite{PDG}.  The 
leading correction in Eq.~(\ref{9}) comes from 
the factor $C_M$ and later on when looking for 
CSV-effects, we will take into account this 
factor only. 

Notice that in the approximation considered above,
the quantity $A_{\pi}$, according to Eq.~(\ref{8}), 
does not depend on scattering angle $\theta$.  
This is the consequence of the simplification we 
used.  Namely, we took into account the impulse 
approximation with the $\pi N$ scattering in the 
$P_{33}$ wave.  As was demonstrated in \cite{ma82}, 
the inclusion of others S- and P-waves does not 
change the picture dramatically but leads to a
smooth dependence of $A_{\pi}$ versus scattering 
angle $\theta$.  (Note, the deviation from 
calculated constant value is much smaller than 
the experimental data.)  Nevertheless, as was 
shown in \cite{ma82}, the inclusion of the CSV 
effect in the form (\ref{8}) already raises the 
possibility of describing the observed CSV on the 
deuteron at 143~MeV for scattering angles $\theta 
\leq 80^{\circ}$.

\item
{\underline {Double-Scattering Approximation}}

The $\pi d$ differential cross section in the 
approximation (\ref{6}) has a minimum at the 
scattering angle around $90^{\circ}$, where the 
non-spin-flip amplitude vanishes.  For this 
reason, the contribution from the 
double-scattering term may be essential in this 
region of scattering angles.  There are three 
diagrams for the double-scattering process which 
are depicted in Fig.~3.  The sum of these 
amplitudes is proportional to the combination
\begin{eqnarray}
\frac{1}{3}~[f_{33} (k)]^2 + \frac{1}{3}~[f_{33} 
(k)]^2 - \frac{2}{9}~[f_{33} (k)]^2,
\label{10} \end{eqnarray}
where the last term comes from the diagram with 
the virtual charge-exchange (Fig.~3c).  To 
estimate the contribution of diagrams of Fig.~3, 
let us use the so-called fixed-centers 
approximation.  This method for $\pi d$ 
scattering was first used by Brueckner 
\cite{br53} (see also ref. \cite{Gibbs}).  Its 
accuracy was later estimated by Kolybasov and 
Kudryavtsev \cite{Kudr1} and \cite{Kudr2}.  

The expression of the double-scattering diagrams 
without elementary $\pi N$ spin-orbit forces in 
this fixed centers approximation has the form 
\cite{Kudr2}:
\begin{eqnarray}
f^{(2)}_{\pi d} = \frac{4}{3}~f_{33} (k)~2~F_2 
( \theta , k) 
\phantom{xxxxxxxxxxxxxxxxxxx} \nonumber \\
= \frac{4}{3}~f_{33} (k)~2~(1 - \frac{1}{3})~ 
f_{33} (k) ~\hat{\vec{k_i}}\cdot\hat{\vec{k'_j}}
\phantom{xxxxxxxxxx} \nonumber \\
\int\Psi _D ^2 (r)~e^{i (\frac{\vec{k} + 
\vec{k'}}{2}) \cdot\vec{r}}~(h_1 (r) ~
\hat{\vec{r_i}}\cdot\hat{\vec{r_j}} + h_2 (r) 
~\delta _{ij} )~d \vec{r}, 
\label{11} \end{eqnarray}
where the functions $h_1 (r)$ and $h_2 (r)$ are
\begin{eqnarray}
h_1 (r) = \frac{e^{i k r}}{r} - \frac{3 e^{i k 
r}}{k^2 r^3} + \frac{3}{k^2 r^3} + \frac{3 i 
e^{i k r}}{k r^2},
\label{12} \end{eqnarray}
\begin{eqnarray}
h_2 (r) = \frac{e^{i k r}}{k^2 r^3} - \frac{1}
{k^2 r^3} - \frac{i e^{i k r}}{k r^2},
\label{13} \end{eqnarray}
and the factor (1 - $\frac{1}{3}$) in the right 
hand side of Eq.~(\ref{11}) is specially 
introduced to clear up the relation between 
relative contributions of the elastic 
double-scattering term (it is proportional to 
$1$) and the virtual charge-exchange diagram 
(it is $\propto - \frac{1}{3}$).

This form of the functions $h_1 (r)$ and $h_2 (r)$ 
corresponds to a certain choice for the off-shell 
dependence for $f_{\pi N}$ amplitudes.  For more 
details see \cite{Kudr2}.  In expression 
(\ref{11}), $\hat{\vec k}$ and $\hat{\vec r}$ are 
the units vectors, $\hat{\vec k} = \vec k /k$, 
$\hat{\vec r} = \vec r /r$, and $\hat k_i$ is the 
$i$-component of this vector.

The sum of the single- and double-scattering 
diagrams in this approximation 
\footnote{We omit temperarily the spin-flip 
amplitudes taking into account only the 
non-spin-flip amplitudes.  The inclusion of 
spin-flip will be done later.} is
\begin{eqnarray}
f_{\pi d}^{(1 + 2)} = \frac{4}{3}~f_{33} (k) 
~ 2 ~ [F_D ( \theta ) ~cos \theta + 
ReF_2 ( \theta ) + i~ ImF_2 ( \theta )].
\label{14} \end{eqnarray}
The functions $F_D ( \theta ) cos \theta$, $ReF_2 
( \theta )$, and $ImF_2 ( \theta )$ are shown in 
Figure~4.  We see from this Figure that the 
amplitude of double-scattering is strongly 
suppressed at forward angles versus 
single-scattering.  But at larger than 
$90^{\circ}$-angles, the contributions of 
single- and double-scattering are comparable.  
Clearly, the inclusion of the interference 
effects at this angular range will be essential.

\item
{\underline {Spin-Flip Amplitude}}

Now, we take into account both the non-spin-flip 
and spin-flip parts of the elementary $\pi 
N$-amplitude~(\ref{3}).  As in our previous 
discussion, we will take into account the single- 
and double-scattering terms without any recoil 
effects (i.~e. in the fixed-center approximation).  
The double-scattering term of the $\pi 
d$-scattering amplitude is
\begin{eqnarray}
f^{(2)}_{\pi d} = \frac{8 \pi \left( E^{\pi N}_{cm} 
\right)^2}{m\,E^{\pi d}_{cm}}\, N\, f^2_{33}(k_{cm}) 
\inqq \,\frac{U}{s^2 - k^2 - i 0}. 
\label{15} \end{eqnarray}
Here $N$ is the isotopic factor, which has been 
already used in Eq.~(\ref{10}), for $\pi^{\pm} 
d$-scattering $N = 4/9 = 1/3 + 1/3 - 2/9$. 

The denominator $\,s^2 - \!k^2\! - i 0$ comes from 
the pion propagator, where $\vec s = \vec k_1 + 
\vec q - \vec q_1$ is the virtual pion 3-momenta 
in the lab. system.  $\,U$ stands for the 
expression which includes the spin effects, 

\begin{eqnarray}
U = Tr\{O' S_2 O S_1\},
\phantom{xx}
O = \frac {1}{\sqrt{2}}\, \vec \epsilon \cdot 
\vec\sigma \, ,
\phantom{xx}
%O' = \frac {1}{\sqrt{2}}\, \vec {\epsilon'} 
%\cdot \vec\sigma \, ,
%\phantom{xx}
S_2 = 2\, \hat {\vec s} \cdot \hat {\vec k'} +
i\vec\sigma\cdot [\hat {\vec s} \times \hat 
{\vec k'} ]\, ,
\label{16} \end{eqnarray}
$$
%S_1 = 2\, \hat {\vec k} \cdot \hat {\vec s} +
%i\vec\sigma\cdot [\hat {\vec k} \times \hat 
%{\vec s} ]\, ,
%\phantom{xxxx}
S_1  = 2\, \hat {\vec k} \cdot \hat {\vec s} +
i\vec\sigma\cdot [\hat {\vec s} \times \hat 
{\vec k} ] . $$
Here  $O$ is spin operator in the S-wave part
of the initial deutron wave function, and $O' = 
(1/\sqrt{2})\, \vec {\epsilon'} \cdot 
\vec\sigma \,$ is the same for the final deutron; 
$\, S_{1,2}$ are spin parts of the $\pi 
N$-amplitudes~\footnote
{
%Here the upper index ``c" denotes the operation $S^c 
%\equiv \sigma_2 \tilde S \sigma_2$ (by this, $1^c = 1$ 
%and $\sigma^c = - \sigma$).  This operation arises in 
%the Feynman diagram technique, when the deutron WF 
%(see the text after Eq.~(\ref{3})) is used in the 
%representation, in which one of the nucleon spinors 
%$w$ is taken in charge-conjugated form $\sigma_2 w^*$.  
%This technique is discussed more thoroughly in the 
%Appendix of the paper \cite{Tar} (see also the reference 
%therein).  The same ``c"-operation is also used, when 
%calculating isospin factors.  The isosinglet WF of the 
%deutron may be written as $(1/\sqrt{2}) 
%\chi^+_2\tau_2\chi^*_1$ ($\chi_{1,2}$ are nucleon 
%isospinors), and the factor $N$ in Eq.~(\ref{15}) is 
%$N = (1/3)^2 \{(2 + \vec t \vec {\tau}^{\,c}\,)(2 + 
%\vec t \vec {\tau})\}$, where $\tau^c \equiv \tau_2 
%\tilde\tau \tau_2 = - \tau$.
The technique  we used is discussed in more details in 
our recent paper \cite{Tar}.};
$\hat {\vec s} = \vec s /s$ is the unit vector.  Let 
us represent $U$ as
\begin{eqnarray}
U = \hat{\vec s}_i \cdot\hat{\vec s}_j ~Q_{ij}
\label{17} \end{eqnarray}
and define the integral
\begin{eqnarray}
I_{ij} = \inqq\,\frac{\hat{\vec s}_i 
\cdot\hat{\vec s}_j}{s^2 - k^2 - i0}. 
\label{18} \end{eqnarray}
The tensor $O_{ij}$ in Eq.~(\ref{17}) can be obtained 
from the Eqs.~(\ref{16}).  The integral~(\ref{18}) 
may be rewritten in the form:
\begin{eqnarray}
I_{ij} = J_1\,\hat{\vec\kappa}_i\cdot\hat{\vec\kappa}_j 
+ J_2 ~\delta_{ij}\, ,
\phantom{xx} {\rm where}
\phantom{xx}
\hat {\vec \kappa}=\vec \kappa /\kappa\, ,
\phantom{xx} 
\phantom{xx}
\vec \kappa = (\vec k + \vec k')/2\, .
\label{19} \end{eqnarray}
Here the quantities $J_1$ and $J_2$ are complex 
functions, which depend on $k$ and $\theta$.  They 
depend on the deutron WF as well, and are given in 
the Appendix.

Using Eqs.~(\ref{17}) and~(\ref{18}), we obtain for 
$f^{(2)}_{\pi d}$ the expression of the type 
$f^{(2)}_{\pi d} \sim I_{ij} Q_{ij}$.  Let us 
rewrite the amplitudes $f^{(1)}_{\pi d}$ and 
$f^{(2)}_{\pi d}$ in the form:
\begin{eqnarray}
f^{(1)}_{\pi d} = A_1 \epsilon_i \epsilon'_j 
T^{(1)}_{ij}\, ,
\phantom{xx}
%f^{(2)}_{\pi d} = A_2 I_{ij} Q_{ij} =
%A_2 \epsilon_i\epsilon'_j T^{(2)}_{ij}\, ,
f^{(2)}_{\pi d} = A_2 \epsilon_i\epsilon'_j 
T^{(2)}_{ij}\, ,
\label{20} \end{eqnarray}
where the tensor $T^{(1)}_{ij}$ can be obtained from 
Eq.~(\ref{5}), and $T^{(2)}_{ij}$ -- from the 
relation $I_{ij} Q_{ij} = \epsilon_i\epsilon'_j 
T^{(2)}_{ij}$.  Finally, we get:
\begin{eqnarray}
T^{(1)}_{ij} = 2 z \delta_{ij} + \hat k'_i \hat k_j - 
\hat k'_j \hat k_i\, ,
\phantom{xx}
T^{(2)}_{ij} = a_{ij} J_1 + b_{ij} J_2\, ,
\phantom{xx} {\rm and}
\phantom{xx}
\label{21} \end{eqnarray}
$$
a_{ij} = \frac{1}{2} (5 + 3 z) \delta_{ij} - 2 \hat 
\kappa _i \hat \kappa _j + 3 \hat k'_i \hat k_j - 
\hat k'_j \hat k_i\, ,
\phantom{xx}
b_{ij} = 4 z \delta_{ij} + 5 \hat k'_i \hat k_j - 3 
\hat k'_j \hat k_i\, ,
$$
where $z = (\hat {\vec k} \cdot \hat {\vec k'})$.
The values $A_1$ and $A_2$ in the Eqs.~(\ref{20}) 
for the case of $\pi^+ d$-scattering are
\begin{eqnarray}
A_1 = \frac{2~(m + \omega)}{2~m + \omega}\,(f_1 + 
\frac{1}{3}f_2)~F_D(\theta)\, ,
\phantom{xxxxxxxxxxxxxxxxxxxxxx}
\label{22} \end{eqnarray}
$$
A_2 = \frac{8 \pi (m + \omega)^2}{m~(2~m + 
\omega )}\, \frac{2}{3}~f_2~(f_1 - \frac{1}{3}~f_2)
\phantom{xx}
\left(f_i = \frac{1}{k_{cm}}\,\frac{\Gamma/2}{w_i - 
w - i\Gamma/2}\right)
$$
(here we use more accurate values $E^{\pi N}_{cm} = 
m + \omega\,$ and $E^{\pi d}_{cm} = 2 m + \omega\,$  
than in the simplificated version used in 
Eq.~(\ref{5})).  In the case of the $\pi^- d$ 
elastic scattering, one should substitute $f_1 \to 
f_4$ and $f_2 \to f_3$ in expressions (\ref{22}).  
If $\Delta$-mass splitting is absent, then 
Eqs.~(\ref{22}) are reduced to:
\begin{eqnarray}
A^{(0)}_1 = \frac{2~(m + \omega)}{2~m + \omega}\,
\frac{4}{3}~f_0~F_D(\theta) ,
\phantom{xxxxxxxxxxxxxxxxxxxxxx}
\label{23} \end{eqnarray}
$$
A^{(0)}_2 = \frac{8 \pi (m + \omega)^2}{m~(2~m + 
\omega )}\, \frac{4}{9}~f_0^2
\phantom{xxxx}
\left(f_0 = \frac{1}{k_{cm}}\,\frac{\Gamma_0/2}{w_0 
- w - i \Gamma_0/2}\right).
$$
%\label{23} \end{eqnarray}

After averaging over initial and summation over 
final polarization of deuteron, we can write the 
final result for the cross section $\sigma (\theta) 
\equiv d\sigma/d\Omega$ as the sum of three terms:
\begin{eqnarray}
\sigma ( \theta ) = \sigma _{11} ( \theta ) + \sigma 
_{12} ( \theta ) + \sigma _{22} ( \theta ) ,
\label{24} \end{eqnarray}
where $\sigma_{11}$ and $\sigma_{22}$ are the 
contributions from the single- and double-scattering, 
respectively, and $\sigma_{12}$ is the single-double 
interference term.  The expressions for these cross 
sections are given below:
\begin{eqnarray}
\sigma_{11}(\theta) = \frac{1}{3}~|A_1|^2 \,
T^{(1)*}_{ij} T^{(1)}_{ij} = \frac{2}{3}~|A_1|^2 \, 
(1 + 5 z^2)\, ,
\phantom{xxxxxxxxxx}
\label{25} \end{eqnarray}
\begin{eqnarray}
\sigma_{12}(\theta) = \frac{2}{3} Re\left[A^* 
_1 A_2 \, T^{(1)*}_{ij} T^{(2)}_{ij}\right] 
\phantom{xxxxxxxxxxxxxxxxxxxxxxx}
\label{26} \end{eqnarray}
$$
\phantom{xx}
= \frac{2}{3} Re\left[A^* _1 A_2\, [(4 + 11 z + 9 
z^2)~J_1 + (8 + 20 z^2)~J_2]\right]\, ,
$$
%\label{26} \end{eqnarray}
%
\begin{eqnarray}
\sigma_{22}(\theta) = \frac {1}{3} |A_2|^2 \, 
T^{(2)*}_{ij} T^{(2)}_{ij} = \frac {1}{3} |A_1|^2 
\, [ \frac {1}{4} (75 + 90 z + 27 z^2) |J_1|^2 
\label{27} \end{eqnarray}
$$
\phantom{xxxxxx}
+ (16 + 25 z + 15 z^2)~(J_1 J^* _2 + J^*_1 J _2) 
+ (34 + 34 z^2) |J_2|^2 ] . 
$$
%\label{27} \end{eqnarray}

Taking in view, that the leading CSV-correction 
comes 
from the mass splitting and this splitting is 
small, it would be useful to represent the 
formula for the cross section in a linearized in 
$\delta m_{\Delta}$ form.  In this limit, the 
expression for asymmetry has the form: 
\begin{eqnarray}
A_{\pi} = - \frac{C_M}{2\sigma^{(0)}\Gamma}
[3 (B_0 + B_0^*)~[ \frac{1}{2}\sigma^{(0)}_{11}(
\theta ) + \sigma ^{(0)}_{22}( \theta )]
\phantom{xxxxxxxxxxxxxxx} \nonumber \\
+ 2 Re [A^*_1 A_2 ~(B_0 + \frac{1}{2} B_0^*)~[(4
+ 11 z + 9 z^2) J_1 + (8 + 20 z^2) J_2]]],   
\label{28} \end{eqnarray}
and correspondingly ratio $r = 1 + 2~A_{\pi}$.  
Here $B_0 = \frac {\Gamma_0 /2}{w_0 - w - i 
\Gamma_0/2}$; the values $\sigma^{(0)}$, 
$\sigma^{(0)}_{11}$ and $\sigma^{(0)}_{22}$ are 
defined by Eqs.~(\ref{24}), ~(\ref{25}), 
and~(\ref{27}), respectively, after substitutions 
$A_1\to A^{(0)}_1$ and $A_2\to A^{(0)}_2$ from 
Eqs.~(\ref{23}).

Hence all the CSV-corrections depend on the same 
linear combination of masses, as in the 
single-scattering term, i.~e. on the parameter
$C_M \simeq 4.6~MeV$.  Note that the inclusion
of the double-scattering introduces no new 
parameters, i.~e. the effect is still primarily 
dominated by $C_M$.

\item
{\underline {Coulomb Interaction}}

Now, we consider the fact that the charged pions 
interact with the deuteron by the Coulomb force.  
The elementary $\pi N$-amplitude, which corresponds 
to the interaction of a pion with a proton via 
$\gamma$-exchange, is drawn in Figure~5.  In terms 
of bi-spinors, the expression for this diagram is 
$$M^{(\gamma)}_{\pi p} = \frac{4 \pi e^2}{t}\bar 
u_2 (k_1 + k_2)_{\mu} \gamma ^{\mu} u_1.$$  
Neglecting the magnetic interaction and adding the 
Coulomb phase, we finally get for the Coulomb 
amplitude 
\begin{eqnarray}
f^{\gamma} = \frac{M^{( \gamma )}_{\pi p}}{8 \pi 
(m + \omega )} = 
\phantom{xxxxxxxxxxxxxxxxxxxxxxxxxxxxxx} \nonumber \\
- \frac{e^2}{2 k^2_{cm} \sin^2 ( 
\frac{\theta}{2} )} \frac{\omega m}{(m + \omega )} 
\exp{\left[- \frac{2 i e^2}{k_{cm}} \frac{\omega 
m}{(m + \omega)}\ln \left( \sin\frac{\theta}{2} 
\right) \right]},
\label{29} \end{eqnarray}
where $e^2 = \frac{1}{137}$.  Below we use the 
amplitude $f^{\gamma}$ (\ref{29}) convoluted 
with the proton density of deuteron as a crude 
approximation to the Coulomb pion-deuteron 
scattering amplitude $f^{(\gamma)}_{\pi d}$.  We 
took into account the square of this amplitude 
as well as its interference with single and 
double scattering terms.  Technically, it is 
more suitable to introduce in addition to the 
values $A_1$ and $A_2$ the new one $A_C$:
\begin{eqnarray}
A_C = \frac{2 (m + \omega )}{2~m + \omega}~(f_1 
+ f^{\gamma} + \frac{1}{3} f_2)~F _D( \theta ).
\label{30} \end{eqnarray}

In terms of these $A_1$, $A_2$, and $A_C$, the 
cross sections $\sigma_{11}$ and $\sigma_{12}$ 
now have the form:
\begin{eqnarray}
\sigma_{11}( \theta ) = \frac{2}{3}~[6 z^2 \mid 
A_C \mid ^2 + (1 - z^2) \mid A_1 \mid ^2], 
\label{31} \end{eqnarray}
\begin{eqnarray}
\sigma _{12}( \theta ) = \frac{2}{3}~Re[A_C^* 
A_2 ~ [(11 z + 13 z^2)~J_1 + 28 z^2 ~J_2] 
\nonumber \\
+ A_1^* A_2 [(4 - 4 z^2)~J_1 + (8 - 8 
z^2)~J_2]],
\phantom{xxxxx}
\label{32} \end{eqnarray}
and the expression for $\sigma_{22}$ is given 
by expression (\ref{27}).

Note, that a fairly thorough study of the Coulomb 
effects on pion-deuteron scattering and CSV 
effects were performed in Ref. \cite{Froh}, see 
also \cite{ma82} and \cite{ma84}.  As we are mainly 
interested in looking for CSV effects, which comes 
from the double scattering term and $\Delta$-isobars 
mass splitting, we limit ourselves to the Coulomb 
amplitude in crude approximations (\ref{29}).  Note 
also, that another source of CSV effects in the $\pi 
d$ elastic scattering may come from the direct 
isospin breaking effect in the strong $\pi 
N$-amplitudes, see in this connection Ref. 
\cite{Gibbs3}.  We do not consider the influence 
of this possible interaction on the value of 
$A_{\pi}$ in this paper.

The curves for asymmetry $A_{\pi}$ with the Coulomb 
interaction taken into account are given in 
Figures~6.  If we consider the $\pi ^- d$ scattering 
instead of $\pi ^+ d$, we should substitute in the 
expressions (\ref{22} and \ref{30}): $f_1 \to f_4$, 
$f_2 \to f_3$, and $f^{\gamma} \to -f^{\gamma}$.  
From Fig.~6, we see that single-scattering does not 
depend on the scattering angle but a change of 
sign of the asymmetry does occur between 180 and 
220~MeV according to the expression, given by 
Eq.~(\ref{9}).
\end{enumerate}

\section{Conclusion and Future Prospects}
\label{sec:conc}

In making comparisons of the experimental data for 
asymmetries (Fig.~1) and the corresponding 
theoretical curves (Figs.~6), we conclude that any 
CSV-effects due to the double-scattering terms are 
indeed very small and within uncertainties of 
experimental data.  Our approach gives indications 
of some enchancement of $A_{\pi}$ in the region of 
angles around 90~degrees.  For example, at 
$T_{\pi}$ = 180~MeV (in a range of maximum effect 
of the Delta resonance) there is evidence for the 
growth of $A_{\pi}$ from $A_{\pi}$ = 0.002 at 
$\theta$ = 50$^{\circ}$ to $A_{\pi}$ = 0.015 at 
$\theta$ = 85$^{\circ}$ (We can expect some 
enhancement at 85$^{\circ}$ due to the behaviour 
of $F_D ( \theta ) cos \theta$, $ReF_2 ( \theta )$, 
and $ImF_2 ( \theta )$ shown in Figure~4.)  But the 
growth of $A_{\pi}$ is not large.  The energy 
behaviour of $A_{\pi}$ at 85$^{\circ}$ is shown on 
Fig.~7.  At the same time, experimental errors for 
asymmetry in this region of angles are the order 
of one percent.  The same is true for other 
energies.  We conclude that to confirm these
theoretical predictions for the asymmetry on 
the deuteron, one needs to have data that are
approximately 2 $-$ 3 times better in
precision than currently available. This does not 
seem to be planned in the near future.

The situation may be quite different in the $^3He/ 
^3H$-case.  There are two arguments as to why one 
may expect the CSV-effect to be larger for these 
nuclei:

\begin{itemize}
\item
The enchancement of effect in $^3He/ ^3H$ case 
in comparison to deuteron may take place because 
of a smaller role of the spin-flip terms in the 
single-scattering approximation.  In this 
approximation for the deuteron case, the ratio of 
non-spin-flip to spin-flip terms in the cross 
section is 6:1.  This ratio is quite a bit larger 
for the $^3He/ ^3H$-case.  So, the role of 
double-scattering terms in the region of angles 
around 90~degrees may be much more pronounced for 
these nuclei. 
\item
The number of double-scattering diagrams also
increases due to the large number of possible  
rescattering combinations.  This further enhances
the role of double-scattering terms in comparison
to the deuteron case.
\end{itemize}

The role of Fermi motion has not been discussed.
This is primarily because the main aim of this 
work has been to investigate processes which 
could possibly reproduce the observed structure 
in $\pi d$ asymmetries.  Fermi motion is expected 
to broaden the ``signal" but not lead to the 
sought-after structures.  Moreover, in the case 
of the deuteron, where the asymmetry signal, 
both observed and calculated, is small, it is 
presumably premature to discuss corrections 
before the magnitude of the effect is reasonably 
understood.

Using the developed formalism on $\pi d$ elastic 
scattering, the $^3He/ ^3H$ case is considered 
in forcoming paper.

%%%%%%%%%%%%%%%%%%%%%%%%%%%%%%%%%%%%%%%%%%%%%%%%%%%%%
\acknowledgments
The authors acknowledge useful communications with 
B.~L. Berman, J. Friar, and S. Kamalov.
One of us (A.~K.) acknowledges the hospitality 
extended by the Center for Nuclear Studies of The 
George Washington University.  This work was 
supported in part by the U.~S.~Department of Energy 
Grants DE--FG02--99ER41110 and DE--FG02--95ER40901 
with the Russian grant for Basic Research 
N~98--02--17618.  I.~S. gratefully acknowledge a 
contract from Jefferson Lab under which this work 
was done.  The Thomas Jefferson National Accelerator 
Facility (Jefferson Lab) is operated by the 
Southeastern Universities Research Association
(SURA) under DOE contract DE--AC05--85--84ER40150.

%%%%%%%%%%%%%%%%%%%%%%%%%%%%%%%%%%%%%%%%%%%%%%%%%%%%%
\section{Appendix}
\label{appen}
Here we give the expressions for the integrals $J_1$ 
and $J_2$.
\begin{eqnarray}
J_1 = \frac{1}{4}\int dr~ r^2 \psi ^2 (r) ~[(3 E_2 - 
E_0)~h_1 (r)],
\phantom{xxxxxxxx} \nonumber \\
J_2 = \frac{1}{4}\int dr~ r^2 \psi ^2 (r) ~[(E_0 - 
E_2)~h_1(r) + 2 E_0~h_2 (r)].
\label{33} \end{eqnarray}
Here $E_n = \int\limits_{-1}^{+1} e^{i \kappa rz} 
z^n dz$, $\kappa = k \cos (\frac{\theta}{2}) = k 
\sqrt{(1+z)/2}$ and functions $h_1 (r)$  and $h_2 
(r)$ are given in the main text, see Eqs.~(\ref{12}) 
and (\ref{13}).

Let us calculate the integral $J_1$. For this 
purpose, it is suitable to use the following 
representation for underintegral function:
\begin{eqnarray}
(3 E_2 - E_0) h_1 (r) = \sum _{m = 1}^{16} a_m 
\frac{e^{i b_mr}}{r^{n_m}}.
\label{34} \end{eqnarray}
Here $\, n_m = 2,~3,~4,~5,~6,~4,~5$, and $6$ for $m 
= 1,~2,~3,~4,~5,~6,~7$, and $8$, respectively; $n_m 
= n_{m - 8}$ for $9 \le m \le 16$, and 
\begin{eqnarray} 
a_1 =  -2 i k^{-1} x^{-1}, 
\phantom{xxxxxxxxxxxxxxx} \nonumber \\
a_2 =   6   k^{-2} x^{-1} (1 + x^{-1}), 
\phantom{xxxxxxxxxx} \nonumber \\
a_3 =   6 i k^{-3} x^{-1} (1 + 3 x^{-1} + x^{-2}),
\phantom{xxx} \nonumber \\
a_4 = -18   k^{-4} x^{-2} (1 + x^{-1}), 
\phantom{xxxxxxx} \nonumber \\
a_5 = -18 i k^{-5} x^{-3}, 
\phantom{xxxxxxxxxxxxxx} \nonumber \\
a_6 =  -6 i k^{-3} x^{-1}, 
\phantom{xxxxxxxxxxxxxxx} \nonumber \\
a_7 =  18   k^{-4} x^{-2}, 
\phantom{xxxxxxxxxxxxxxxx} \nonumber \\
a_8 =  18 i k^{-5} x^{-3}, 
\phantom{xxxxxxxxxxxxxxx} \nonumber \\
b_1 = b_2 = b_3 = b_4 = b_5= (1 + x)~k, 
\phantom{x} \nonumber \\
b_6 = b_7 = b_8 = x~k, \phantom{xxxxxxxxxxxxx}
\label{35} \end{eqnarray}
where $x = \cos ( \frac{\theta}{2})$.  These 
Eqs.~(\ref{28}) after the replacement $x \to 
-x$ define the values $a_m$ and $b_m$ for $9 
\le m \le 16$ as $a_m = a_{m - 8}$ and $b_m = 
b_{m - 8}$.

In calculations, we use a realistic deuteron 
wave function (in $S$-wave approximation) of 
the Bonn potential \cite{Bonn}, parametrized 
as $\psi (r) = \sum _{i} c_i \frac{e^{- 
\alpha _i r}}{r}$, where $\alpha _i > 0$.  
With this form of $\psi (r)$, we get
\begin{eqnarray}
J_1 = \frac{1}{4}\int\sum _{ijm} c_i c_j 
a_m e^{(i b_m - \alpha _i - \alpha _j) r} 
\frac{dr}{r^{n_m}} .
\label{36} \end{eqnarray}
To evaluate this integral, one may use a general 
relation
\begin{eqnarray}
\int_0^{\infty}\sum _{i} c_i e^{a_i x} 
\frac{dx}{x^{n_i}} = \sum _{i} c_i \frac{
a_i^{n_i - 1}}{(n_i - 1)!} [S_{n_i - 1} 
- \ln a_i ],
\label{37} \end{eqnarray}
where $S_n = \sum _{k = 1}^n \frac{1}{k}$ and 
$S_0 = 0$.  The formula (\ref{27}) is derived 
for the case $n_i \geq 1$ and is valid if this 
integral converges (i.~e. $Re\, a_i < 0$ and the 
underintegral function is finite at $x \to 0$). 
These conditions are satisfied for the integral 
(\ref{26}), and we finally get:
\begin{eqnarray}
J_1 = \frac{1}{4}\sum _{ijm} c_i c_j a_m \frac 
{(ib_m - \alpha _i - \alpha _j)^{n_m - 1}}
{(n_m - 1)!}~(S_{n_m - 1} - \ln \sqrt{( \alpha
_i + \alpha _j)^2 + b^2_m} + i~ atan \frac
{b_m}{\alpha _i + \alpha _j}).
\label{38} \end{eqnarray}
To obtain the expression for $J_2$, one may use 
the analogous representation
\begin{eqnarray}
(E_0 - E_2)~h_1 (r) + 2 E_0~h_2 (r) = \sum 
_{m = 1}^{14} a_m \frac{e^{i b_mr}}{r^{n_m}}.
\label{39} \end{eqnarray}

Here $\, n_m = 3,~4,~5,~6,~4,~5$, and $6$ 
for $m = 1,~2,~3,~4,~5,~6$, and $7$,
respectively; $n_m = n_{m - 7}$ for $8\le m\le 
14$ and
\begin{eqnarray} 
a_1 = -2 k^{-2} x^{-1} (1 + x^{-1}), 
\phantom{xxxxxxxxxxxx} \nonumber \\
a_2 = -2 i k^{-3} x^{-1} (1 + 3 x^{-1} +x^{-2}), 
\phantom{xxxxx} \nonumber \\
a_3 =  6 k^{-4} x^{-2} (1 + x^{-1}),
\phantom{xxxxxxxxxxxxx} \nonumber \\
a_4 = 6 i k^{-5} x^{-3}, 
\phantom{xxxxxxxxxxxxxxxxxxxx} \nonumber \\
a_5 =  2 i k^{-3} x^{-1}, 
\phantom{xxxxxxxxxxxxxxxxxxxx} \nonumber \\
a_6 = -6 k^{-4} x^{-2}, 
\phantom{xxxxxxxxxxxxxxxxxxx} \nonumber \\
a_7 = -6 i k^{-5} x^{-3}, 
\phantom{xxxxxxxxxxxxxxxxxx} \nonumber \\
b_1 = b_2 = b_3 = b_4 = (1+x)~k, 
\phantom{xxxxxxxx} \nonumber \\
b_5 = b_6 = b_7 = x~k, \phantom{xxxxxxxxxxxxxxxx}
\label{40} \end{eqnarray}
where $x = \cos ( \frac{\theta}{2}).$
These Eqs.~(\ref{33}) after the replacement $x 
\to -x$ define the values $a_m$ and $b_m$ for 
$8\le m\le 14$ as $a_m = a_{m - 7}$ and $b_m = 
b_{m - 7}$.  Thus, for the integral $J_2$ we get 
the similar Eqs.~(\ref{28}) in which the values 
$n_m$, $a_m$, and $b_m$ are defined by 
Eqs.~(\ref{29}) and (\ref{30}). 

\eject
%%%%%%%%%%%%%%%%%%%%%%%%%%%%%%%%%%%%%%%%%%%%%%%%%%%%%%%%%%
%%%                      References
%%%%%%%%%%%%%%%%%%%%%%%%%%%%%%%%%%%%%%%%%%%%%%%%%%%%%%%%%%

\eject
%%%%%%%%%%%%%%%%%%%%%%%%%%%%%%%%%%%%%%%%%%%%%%%%%%%%%%%%%
%%%                      Figure captions
%%%%%%%%%%%%%%%%%%%%%%%%%%%%%%%%%%%%%%%%%%%%%%%%%%%%%%%%%
{\Large\bf Figure captions} \\
\newcounter{fig}
\begin{list}
{Figure \arabic{fig}.}
{\usecounter{fig}\setlength{\rightmargin}{\leftmargin}}
%%%%%%%%%%% fig. 1
\item
{Asymmetry $A_{\pi}$ at different energies.
(a)  30~MeV, (b)  50~MeV, (c)  65~MeV,
(d) 143~MeV, (e) 180~MeV, (f) 220~MeV,
(g) 256~MeV, and (h) 417~MeV for $\pi d$ elastic 
scattering.  Experimental data are from
\protect\cite{ko93} (open circles),
\protect\cite{ko91} (open triangles),
\protect\cite{ma82} (filled triangles),
\protect\cite{ne90} (filled circles),
\protect\cite{ho79} (open diamonds),
\protect\cite{sm88} (stars),
\protect\cite{ma84} (filled squares), and
\protect\cite{gr95} (filled diamonds).}
%%%%%%%%%%%%%%%%%% fig. 2
\item
{Single-scattering amplitudes for $\pi^+ d$ on 
the proton (a) and the neutron (b).}
%%%%%%%%%%% fig. 3
\item
{Double-scattering amplitudes for $\pi^+ d$ : 
elastic (a) and (b), and with virtual 
charge-exchange (c).} 
%%%%%%%%%%% fig. 4
\item
{Amplitudes for $\pi d$ elastic scattering without 
spin-flip at 140~MeV.  Solid curve gives $F_D 
( \theta ) cos \theta$.  The real (imaginary) parts 
of amplitude $F_2( \theta)$ is plotted with 
dash-dotted (dashed) lines.}
%%%%%%%%%%% fig. 5
\item
{Feynman diagrams for the Coulomb $\pi p$ and 
$\pi d$ amplitudes.}
%%%%%%%%%%% fig. 6
\item
{Asymmetry for $\pi d$ elastic scattering with the
Coulomb interaction taken into account.
(a) 143~MeV, (b) 180~MeV, (c) 220~MeV,
and (d) 256~MeV.
Experimental data are from \protect\cite{ho79}
$-$ \protect\cite{ma84}, \protect\cite{ko91}
$-$ \protect\cite{ne90}.  Notation is the same
as in Fig.~1.  Solid curves give the total
amplitude.  Single (and double) scattering
without Coulomb corrections is shown by
dashed (dash-dotted)curves.}
%%%%%%%%%%% fig. 7
\item
{85$^{\circ}$ energy dependence of asymmetry 
$A_{\pi}$ for $\pi d$ elastic scattering with 
the Coulomb interaction taken into account.  
Notation is the same as is in Fig.~6.}
\end{list}
%%%%%%%%%%%%%%%%%%%%%1
\begin{figure}[tb!]
\centerline{\psfig{file=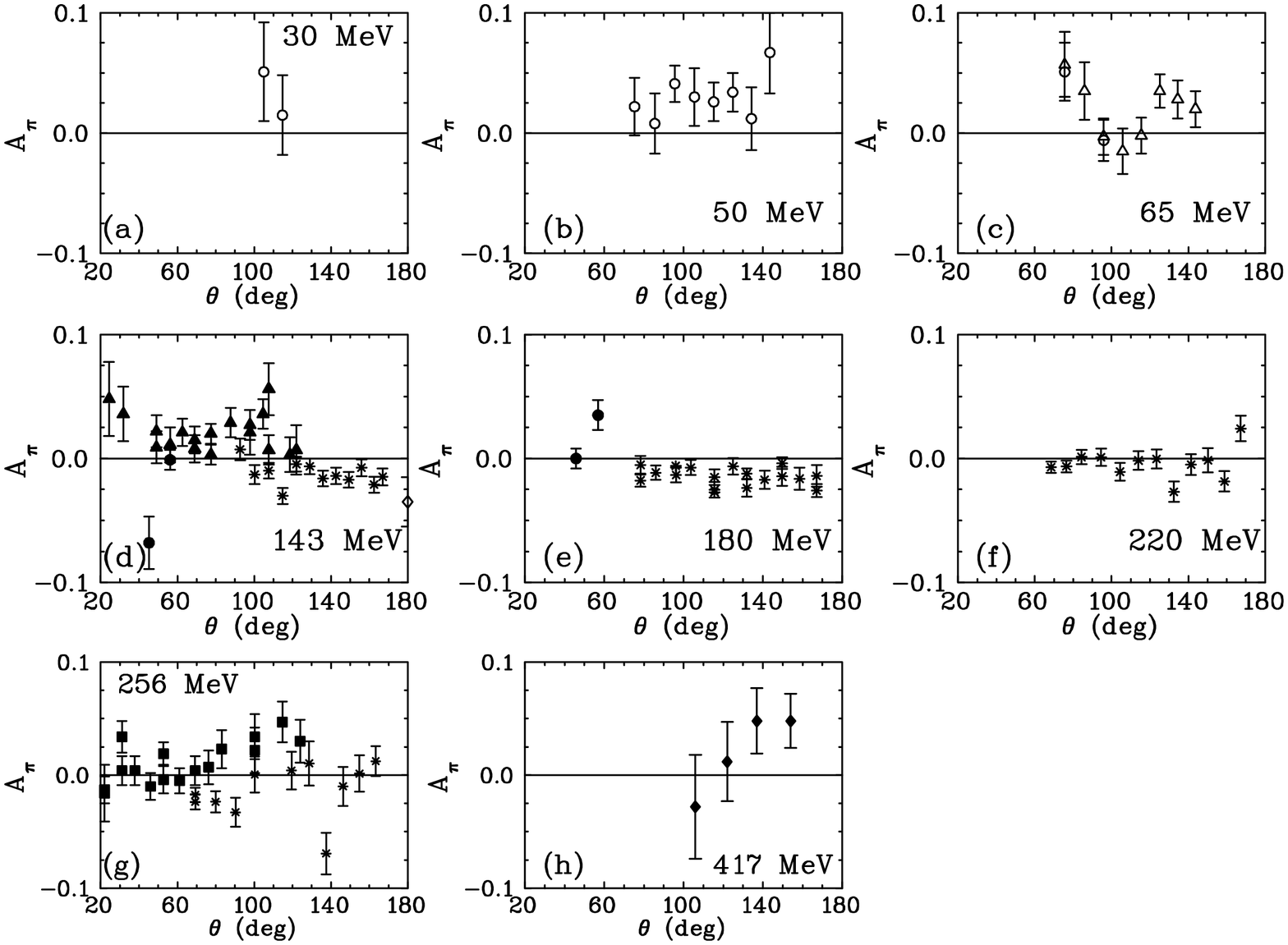,width=.9\textwidth,clip=,silent=,angle=90}}
\caption[fig1]{\label{fig1}}
\end{figure}
%%%%%%%%%%%%%%%%%%%%%2
\begin{figure}[tb!]
\centerline{\psfig{file=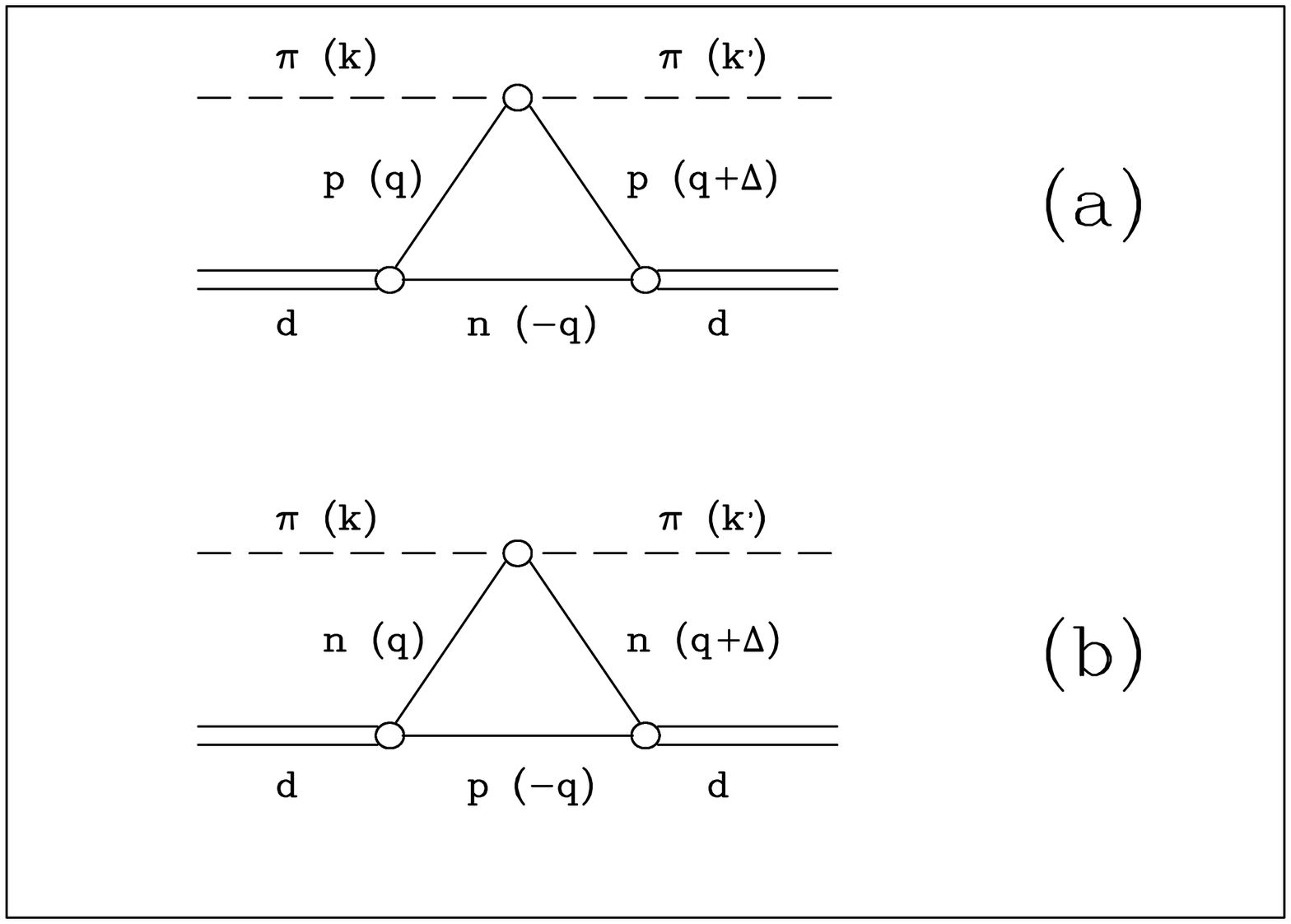,width=.7\textwidth,clip=,silent=,angle=90}}
\caption[fig2]{\label{fig2}}
\end{figure}
%%%%%%%%%%%%%%%%%%%%%3
\begin{figure}[tb!]
\centerline{\psfig{file=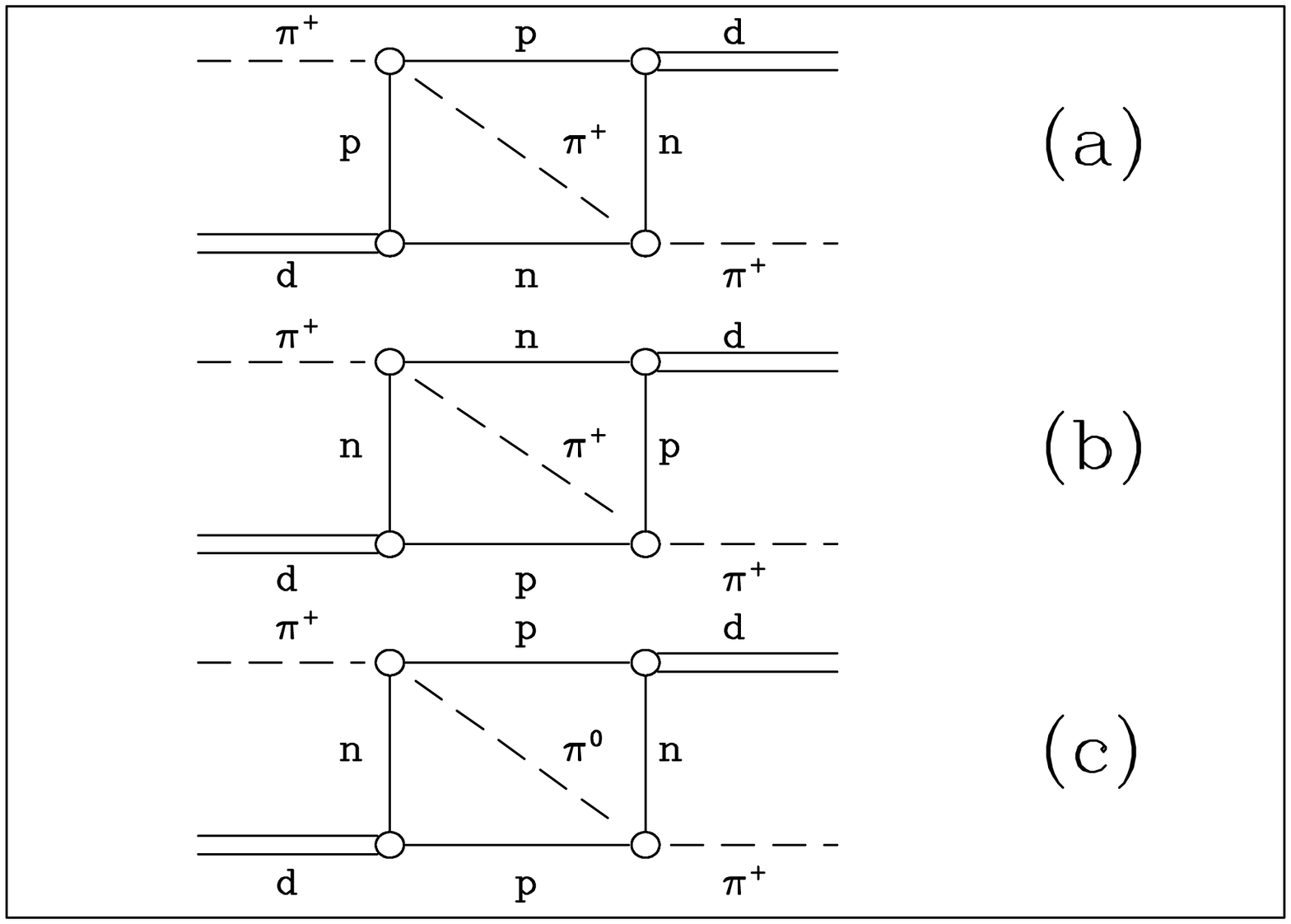,width=.8\textwidth,clip=,silent=,angle=90}}
\caption[fig3]{\label{fig3}}
\end{figure}
%%%%%%%%%%%%%%%%%%%%%4
\begin{figure}[tb!]
\centerline{\psfig{file=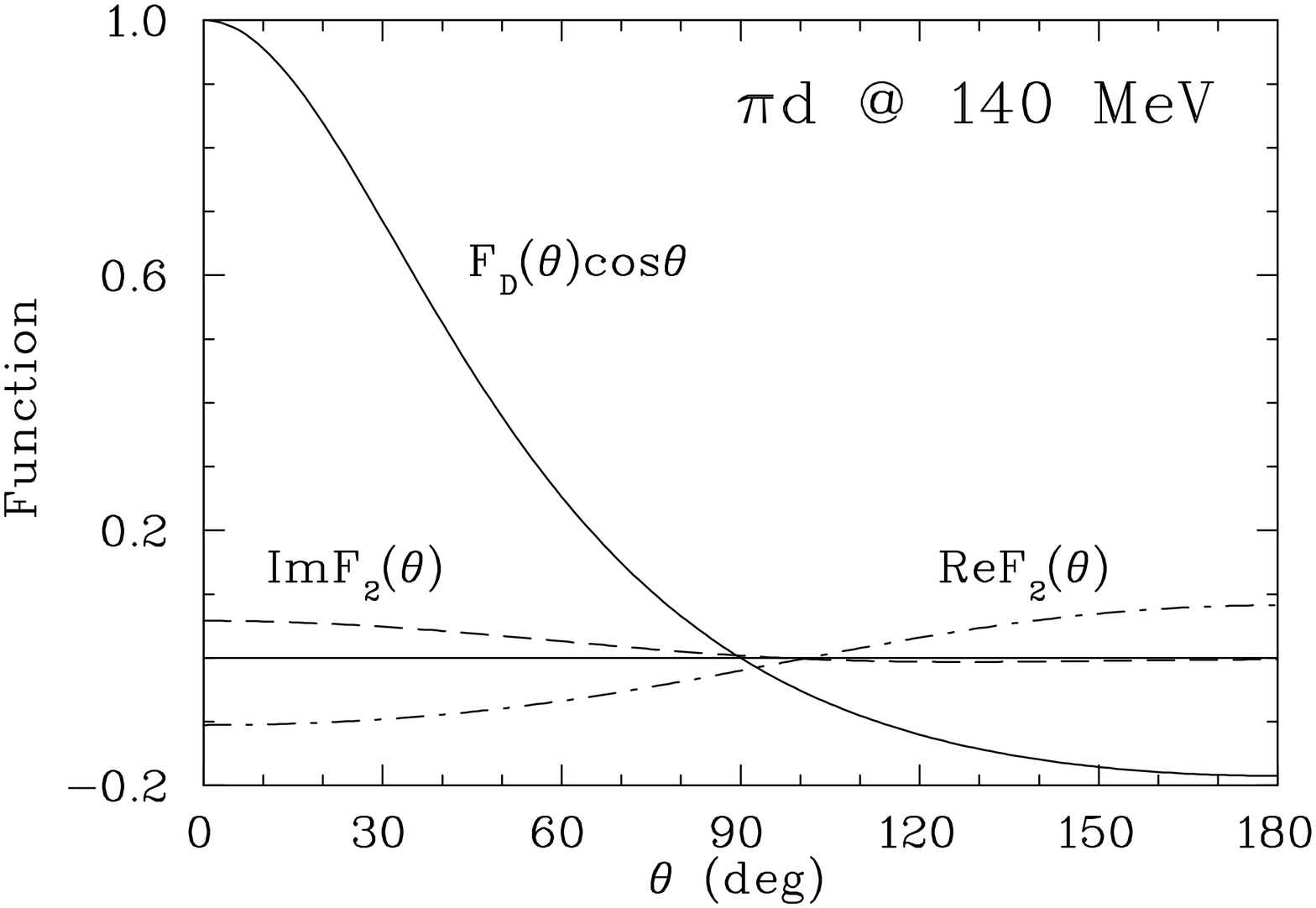,width=.8\textwidth,clip=,silent=,angle=90}}
\caption[fig4]{\label{fig4}}
\end{figure}
%%%%%%%%%%%%%%%%%%%%%5
\begin{figure}[tb!]
\centerline{\psfig{file=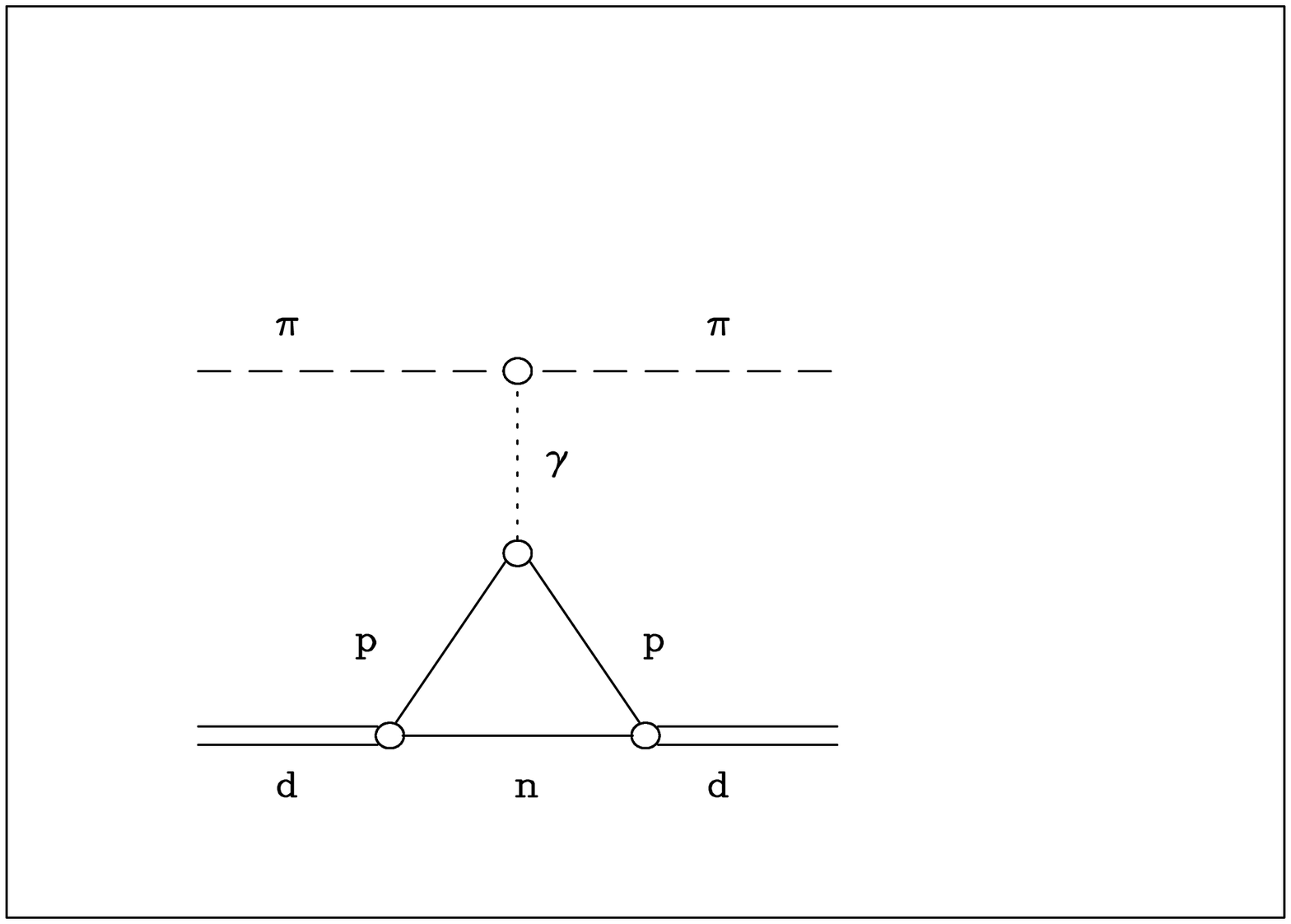,width=.8\textwidth,clip=,silent=,angle=90}}
\caption[fig5]{\label{fig5}}
\end{figure}
%%%%%%%%%%%%%%%%%%%%%6
\begin{figure}[tb!]
\centerline{\psfig{file=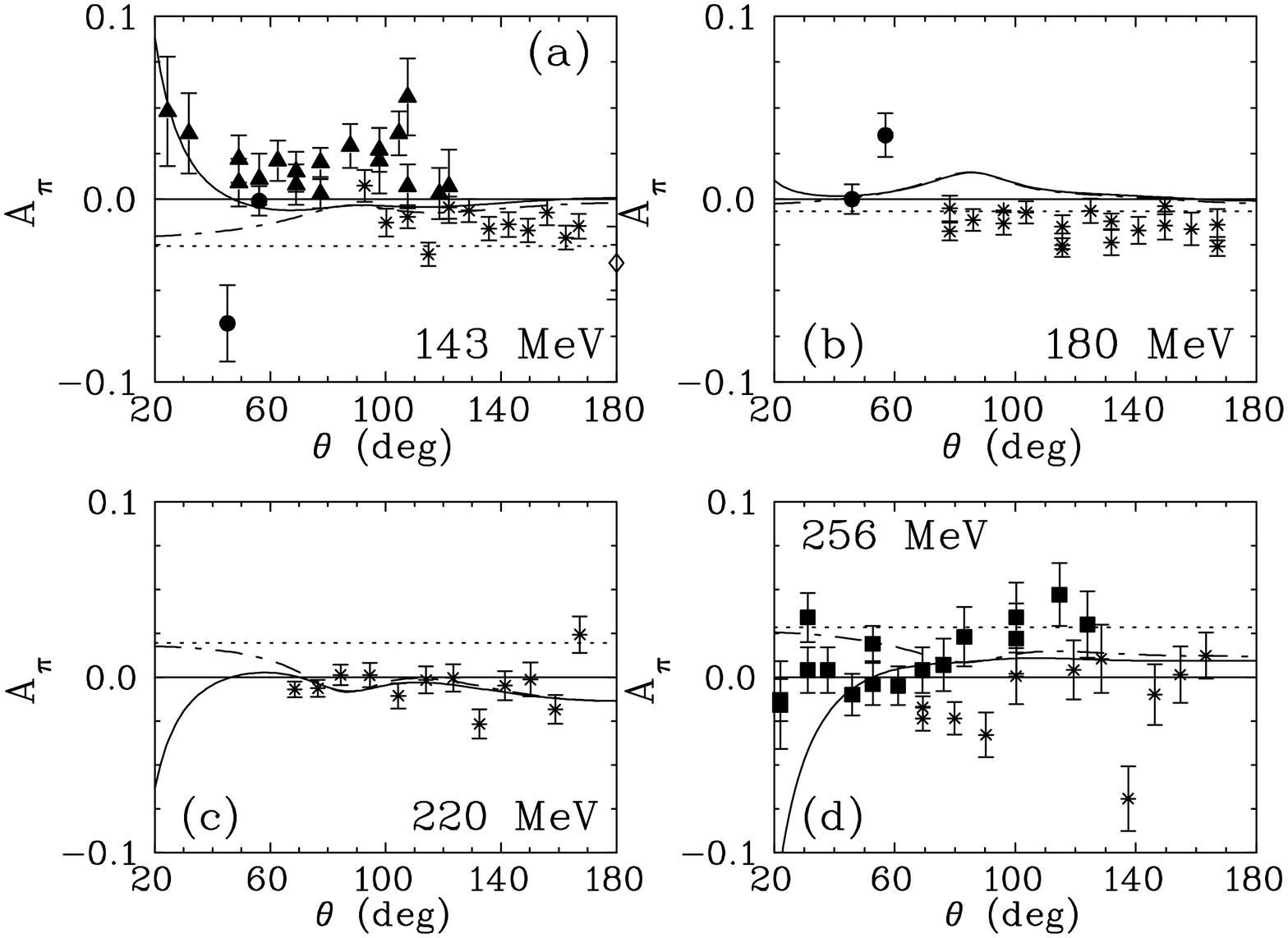,width=.9\textwidth,clip=,silent=,angle=90}}
\caption[fig6]{\label{fig6}}
\end{figure}
%%%%%%%%%%%%%%%%%%%%%7
\begin{figure}[tb!]
\centerline{\psfig{file=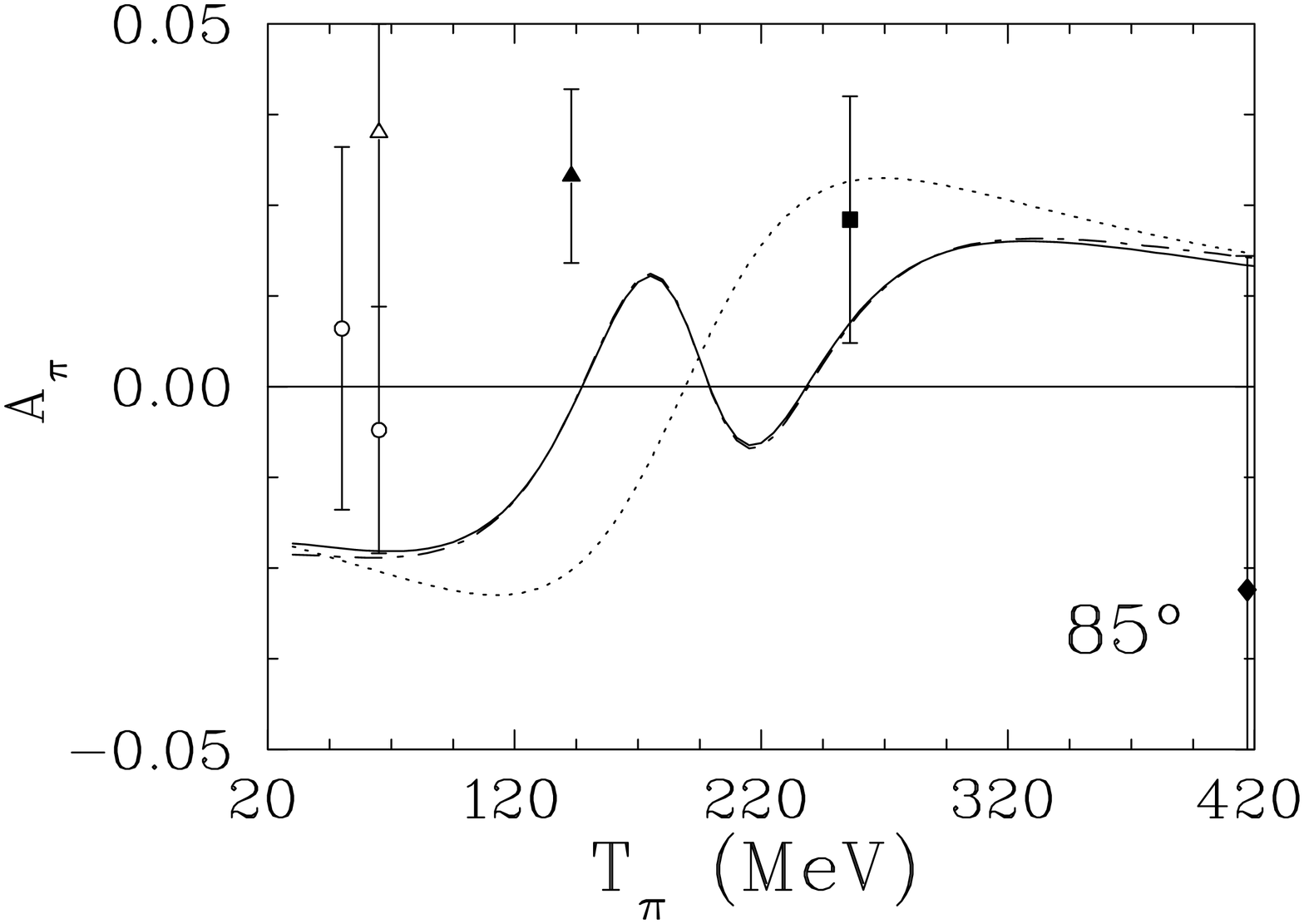,width=.9\textwidth,clip=,silent=,angle=90}}
\caption[fig7]{\label{fig7}}
\end{figure}
\end{document}